\documentclass{article}

\title{Form Follows Function---Do algorithms and applications challenge or drag
behind the hardware evolution?}

\author{Tobias Weinzierl\\School of Engineering and Computing Sciences\\Durham
University, GREAT BRITAIN\\tobias.weinzierl@durham.ac.uk}

\begin{document}

\maketitle

\begin{abstract}
We summarise some of the key statements made at the workshop {\it Form
Follows Function} at ISC High Performance 2016. 
The summary highlights what type of co-design the presented projects
experience; often in the absence of an explicit co-design
agenda.
Their software development picks up hardware trends but it also influences the
hardware development.
Observations illustrate that this cycle not always is optimal for both
sides as it is not proactively steered.
Key statements characterise ideas how it might be possible to integrate both
hardware and software creation closer to the
best of both worlds---again even without classic co-design in mind where new
pieces of hardware are created.
The workshop finally identified three development idioms that might
help to improve software and system design with respect to emerging hardware.
\end{abstract}

\noindent
We summarise some of the key statements made at the workshop {\it Form
Follows Function} held as a half day event at ISC High Performance 2016. 
The workshop organised together with Michael Bader (Technische
Universit\"at M\"unchen, TUM) provided a platform to seven European
supercomputing projects to share their view on the interplay of hardware and
software evolution. 
Invited speakers were 
\begin{itemize}
  \item Jack Dongarra (The University of Manchester), 
  \item Rapha\"{e}l L\'eger (INRIA), 
  \item Peter Messmer (NVIDIA), 
  \item Mark Parsons (Edinburgh Parallel Computing Centre), 
  \item Marie-Christine Sawley (Intel), 
  \item Philipp Schlatter (Royal Institute of Technology, KTH) and 
  \item Xavier Vigouroux (Atos Bull).
\end{itemize}

The motivation to organise the workshop arose from authoring the H2020
project {\em ExaHyPE---An exascale hyperbolic PDE engine} \cite{ExaHyPE} where
hardware trends play an important role, as we hope to write a software
well-suited for future machines yet to be developed.
We expect other projects in similar funding streams
caught between hardware and software evolution, too.
It is reasonable to ask
how they plan to react to new developments as they are predicted in exascale
roadmaps \cite{ExascaleRoadmap} and whether they hope to have in turn an
influence on the hardware realisation.

The roadmaps are dominated by predictions on hardware. 
At the same time, hardware-software co-design is a frequently cited phrase. 
It suggests that software development can have an impact on the hardware
evolution.
It can actively shape.
The workshop members clarified in their talks to which degree this assumption
holds in the context of their projects, what the interaction of hardware
and software development looks like and weather the interplay is positive and
should be fostered or manipulative and slows down scientific progress?
% Notably, we asked the speakers:
% How do predictions on new hardware features impact the projects'
% research agenda? 
% Do statements on hardware-aware algorithm development and hardware-software
% co-design affect particular machine aspects, or is notably the last term a buzzword?  
% To which degree can simulation codes have an impact on what machines are
% designed?

%\noindent
A text such as the present short juxtaposing has to have shortcomings.
It comprises a selection from the given tasks. 
By no means, it is a comprehensive summary of them.
Statements have been chosen that are well-suited to complement each other and
sound interesting to the author.
The arrangement and choice reflect the author's, not the speakers' view.
With a rearrangement and excerpt of statements, context is missing and statements
might be too compacted.
As the workshop invited European projects, this document has a strong European
flavour.
This is important to keep in mind given that we discuss aspects of
co-design---in a business that is dominated by US vendors.
Furthermore, almost all invited projects emphasise aspects of simulation
software development and integration into classic simulation workflows.
We do not really discuss co-design in a co-design setting:
all statements on co-design are made from a scientific computing's software
point of view.
% The academic notion of co-design is beyond scope. 
Last but not least, some statements are on purpose pointed.

\paragraph*{Running in circles: Does co-design happen (outside co-design
projects)?}

%
% Wie schaut er aus
%
Any discussion on hardware-software/software-hardware influence
has to start from a clarification whether such a cycle does exist and
what it looks like.
The workshop opened with a presentation by Jack Dongarra who sketched such a cycle.
LINPACK \cite{LINPACK} with its emphasis on vectors fits to a particular type of
machine.
It was written at a time when it had been important to tackle the thorny fact
that floating point operations are expensive.
LAPACK \cite{LAPACK} anticipates the advent of caches where keeping the floating
point units busy gains importance.
ScaLAPACK's \cite{ScaLAPACK} design was kicked off by multi-node machines with MPI,
while the dusk of BSP triggers the development of Magma \cite{Magma} and Plasma
\cite{Plasma} .
The latter are subject of study in the NLAFET project
\cite{NLAFET}.
Mark Parsons gave another
example as he outlined how the availability of 3D XPoint non-volatile memory
\cite{3DXPoint} laid the foundations of the NEXTGenIO project \cite{NextGenIO}
studying how to use additional memory layers between main memory and hard disk.

While it is easy to follow how hardware development triggers new
algorithmic work---our own ExaHyPE \cite{ExaHyPE} project hypothesising that hardware will suffer from
severe performance fluctuations is an example for this, too---Jack pointed out
that the (Top 500) benchmarks in turn grew downstreamingly into a directing role
for the hardware evolution, as they make vendors tune their machines
towards these benchmarks; though this has never been the intention behind
them in the first place as he emphasised.
Other examples for the influence back are the increasing IO demands of today's 
software as sketched before, or GPGPU modifications as Peter
Messmer illustrated at hands of the Escape project \cite{Escape}:
atomics and double precision would  not have made it into GPUs that fast if
there had not been a demand of these features from the scientific computing side.
After all, machines are procured because of scientific software needs.
{\it So while we see software written from scratch around every ten years
because of transformative hardware developments, in-between software continuously
influences the hardware evolution; mainly by acting as benchmarks or as they
escalate bottlenecks}.

\paragraph*{The Sorcerer's Apprentice: Is it good the way it is?}

Most workshop participants were sceptical whether the cycle of influence 
is a good one the way we experience it right now:
\emph{It orbits around weaknesses and demands.
It is backward looking.}
Mark articulated that he is worried that the evolution even does not take 
the well-known Amdahl numbers into account \cite{Amdahl}:
``I believe strongly in co-design but it happens extremely rarely''.
% He then gave a few examples from previous projects in the QCD context where it
% did work out.

%
% Was stoert also; 
%
In general, benchmarks, i.e.~software, to some degree prescribe what type of machines
are bought.
But then the ``real'' software struggles to use those machines---as they tune
towards few hardware characteristics or can meet certain criteria (peak
performance) only by introducing strong constraints somewhere else.
Jack's linear algebra might be the most popular materialisation of this.
%cross-fertilisation with a strong unidirectional flavour.
Yet, Philipp Schlatter pointed out
that it is not hard to find further examples:
a focus on higher order methods in computational fluid dynamics as studied in the ExaFLOW project
\cite{Exaflow} becomes popular as hardware evolution makes flops
comparably free and vendors widen the vector registers.
Side effects then are challenges such as memory starvation or resiliency if we
increase the clock rate.
Xavier Vigouroux from ESCAPE \cite{Escape} also illustrated
how hardware  influences the applications at Atos Bull and made the audience
take a broader view:
Deep learning has been a hot topic at ISC \cite{ISC}, and deep learning relies
heavily on dense matrix-matrix and matrix-vector multiplications.
Honi soit qui mal y pense. 
Hardware vendors and computing centres have to cheer to see the rise of an
application type that fits to the hardware characteristics.
%We may continue on the beaten track.

%
%
%
{\it More pessimistic fellows might argue that the hardware evolution has become
independent and out of control---a Sourcerer's Apprentice---as it guides 
research directions.} 
Shouldn't scientific questions be there in the front row
irrespective of what hardware is available and irrespective of the fact that, yes, it is great if
new hardware opens new horizons?
Maybe, we do not have co-design, though there is a cycle of
influencing each other and it always has been there.
Maybe, we have co-not-design.

\paragraph*{It's a long way to the top \ldots}

A significant time of the workshop was spent on the question ``Why is it the way
it is?''.
It was Xavier who clarified that one might fall short if speaking in terms of
hardware-influences-software or vice versa. 
He advocated for use cases where we either have a unidirectional influence of a
scientific question onto hardware or software or impact on both with 
real interaction between the two of them.
He then raised two important questions: ``What do we
want to invest for performance of a Use Case'' and ``how specific do we want to
go for performance of a Use Case''?
Economic reasoning seems to be important: hardware architects of
major computer components can and will not tailor their kit specifically to
comparably few simulation software requirements. 
It is thus natural that we see impact from hardware on software more often.
Reiterating Peter's aforementioned remark on GPGPU architectures, it might be
important to admit that HPC as a business is too weak as driving force.
The development of graphics hardware after all had always been
pushed by games and new features well-suited for simulations then were realised
if and only if it is comparably cheap to do so.
Peter continued as he highlighted that Escape
thinks about using computer components that can, for example, solve FFT almost
for free---though again such a specialised development might be driven by other
disciplines---which is then a starting point to
redesign software radically for specialised hardware.

Xavier put this fatalistic point of view into perspective. 
While component manufacturers have to exploit the economy of scales, integrators
have more freedom and can react to user demands comparably flexible.
He advocated for a more intense, earlier conversation with integrators:
{\em Tailor the assembly of hardware components to your needs rather than wait
for new hardware building blocks---which may or may not happen.}
Take back control!

Besides the economic reasoning, the role of benchmarks has to be outlined.
Jack didn't tire to emphasise that he very much would appreciate if more
complex, realistic benchmarks would play a more
important role.
Marie-Christine Sawley diversified the term benchmark.
For this, she gave examples from the Exa2CT project \cite{Exa2CT} as well as
from Intel's in-house strategy.
She pointed out that Intel is used to 
run pathfinding labs to understand very early in the design process
how software might pick up new hardware trends.
Such feedback can be fed back into the development cycle though the latency
there is high.
For pathfinding, it is reasonable not to rely on mature benchmarks but to write
proto-applications/reproducer codes---cooked down proxies that 
identify bottlenecks rather than provide mature functionality, highlight
characteristics of future codes, and allow for a smoother transition to a new architecture when we assemble
systems as well.

Besides benchmarks and proto-applications, Mark added that it might, for
transformative hardware changes, be better to abstract from actual
sources completely and to rely on simulators with software and hardware models
instead.
While projects such as NEXTGenIO can work with models and calibrate these
models already today with realistic data, he however pointed out that we
might fall short to focus on the software only. 
The whole software stack (profilers, e.g.) has to participate in such a co-design.
Irrelevant of the favoured materialisation, there had been broad
agreement on ``the sooner the more software for new hardware the better''.
It is not fair to blame vendors and refer to economic side conditions if the
software lacks behind.
It is important to actively shape the future.

\paragraph*{\ldots if you wanna rock 'n roll: Form follows function}

Starting from Peter's emphasis on the fact that there are various metrics to
consider---time-to-solution, Joules per solution, Watt per solution---three fundamental remarks closed the workshop.

First, there had been broad agreement that a sole emphasis on flops is
insufficient.
Two projects---Philipp's ExaFLOW and our own ExaHyPE---in the workshop rely
heavily on higher order techniques to increase the arithmetic intensity,
i.e.~they tackle the almost-freeness of flops with a particular mathematical
technique.
Michael raised the idea that this will be a trend for other hardware aspects as
well and that {\em we will see more and more XYZ-oblivious algorithms (replace
XYZ with arithmetic intensity or cache today, but there might be other
metrics})---though the term ``oblivious'' was not well-liked by everybody.
He continued to ask whether it makes sense to, for the assessment of new 
architectures as well as algorithmic ideas, rely on studies on strong
scaling, weak scaling and scaling in arithmetic intensity or data structure
regularity in future supercomputing papers---after all, it seems that 
scaling becomes simpler if we increase the data structure regularity
or arithmetic intensity arbitrarily and agnostic of the sense of meaning from an
application point of view.

Second, multiple participants claimed that co-design already on the software
side only is underdeveloped. 
Mark: Does it make sense to continue to translate mathematics into matrices and
vectors in the era of exascale parallelism?
Peter: Can we continue to accept
that the classic development workflow changes all phases of scientific code
development besides modelling and mathematics whenever a new hardware generation
arises?
{\em There seems to be a need to put all phases of scientific computing,
including the mathematical modelling plus hardware choices, to the test}.
All steps in simulations explode (input data size, mesh size, compute
complexity), so it might be insufficient to just port applications, wait for new
hardware generations to trigger action,  and to stick to mathematical paradigms
that have been around 20 years ago.

Last, several talks meandered around the problem of insufficient vector unit
utilisation of particular architectures.
Rapha\"{e}l L\'eger illustrated at hands of the DEEP-ER
\cite{Deeper} that future machines might be highly inhomogeneous and not all code parts might be well-suited for
all machine parts.
So why should a code not wander over a machine and use, at different times,
those parts of the machine that suit its needs?
IO-intense algorithm phases by definition have other requirements than
compute-bound kernels.
Or, the other way round: why should we continue to strive for an
efficient vector unit usage and try, with all our effort, to ensure that a vector unit
does not wait for data? 
It might be more reasonable to {\em build machines and deploy
software such that vector units are always available when they are needed}.

\subsection*{Acknowledgements}

Tobias appreciates the support received from the European Union’s Horizon 2020
research and innovation programme under grant agreement No 671698 (ExaHyPE).
Additional workshop information and slides are available from \cite{Workshop}.

\end{document}